\documentclass{iopart}

\pdfoutput=1

\usepackage{color}
\usepackage{graphicx,epsfig}
\usepackage{bm}
\usepackage{amssymb}
\usepackage{braket}

\def\be{\begin{equation}}
\def\ee{\end{equation}}

\def\bea{\begin{eqnarray}}
\def\eea{\end{eqnarray}}

\def\s{\sigma}
\def\th{\theta}
\def\l{\lambda}
\def\sgn{{\rm sgn}}

\def\nn{\nonumber\\}

\begin{document}
\title[Parity effects in the entanglement of
 XX chains with open boundary conditions] 
{Universal parity effects in the entanglement entropy of
 XX chains with open boundary conditions}
\author{Maurizio Fagotti and Pasquale Calabrese}
\address{Dipartimento di Fisica dell'Universit\`a di Pisa 
and INFN, Pisa, Italy }

\date{\today}

\begin{abstract}
We consider the R\'enyi entanglement entropies in the one-dimensional  XX spin-chains with open boundary conditions in the presence 
of a magnetic field. 
In the case of a semi-infinite system and a block starting from the boundary, 
we derive rigorously the asymptotic behavior for large block sizes  on the basis of a recent mathematical theorem for 
the determinant of Toeplitz plus Hankel matrices. 
We conjecture a generalized Fisher-Hartwig form for the corrections to the asymptotic behavior of this determinant that allows 
the exact characterization of the corrections to the scaling at order $o(\ell^{-1})$ for any $n$. 
By combining these results with conformal field theory arguments, we derive exact expressions also in finite chains with open 
boundary conditions and in the case when the block is detached from the boundary. 

\end{abstract}

\maketitle
\section{Introduction}

The interest in quantifying the entanglement in the ground state of extended quantum systems has risen 
sharply in the last decade \cite{revs}. Remarkably, the results of many investigations 
allowed a deeper and  more precise characterization of many-body systems. 
Furthermore, many surprising connections between fields and techniques apparently disconnected emerged. 

In this paper we will consider the R\'enyi entanglement  entropies in the XX chain with open boundary conditions (OBC). 
As we will review, the leading asymptotic behavior of the entanglement entropy can be deduced directly from known results 
in conformal field theory (CFT) joined with  available exact calculations for the chain with periodic 
boundary conditions (PBC). However, the open chain presents subleading oscillatory corrections to the scaling 
whose first observation dates back to 2006 \cite{lsca-06} and that until now resisted to an analytic computation. 
This study 
 provides a new and unexpected by-product that, 
for some aspects, is even more interesting that the main result itself. 
Indeed, in order to arrive to an analytic result for the entanglement entropy in systems with OBC, we faced the 
problem of calculating leading and subleading behavior of the determinants of matrices that in mathematical 
literature are known as Toeplitz plus Hankel (i.e. they are composed of a part that depends only on the difference between 
row and column indexes and another depending only on their sum). 
Toeplitz matrices have a very long history, culminating with the Fisher-Hartwig (FH) conjecture \cite{fh-c}. 
This conjecture has been proved (in some particular cases) only many years after its formulation by Basor \cite{bm-94}. 
The interest in the corrections to this formula leaded to a generalization
known as generalized FH conjecture \cite{bt-91} that has not yet been proved. This formula has been fundamental to 
provide the corrections to the scaling for the entanglement entropy in systems with PBC 
\cite{ccen-10,ce-10}. 
When moving from PBC to OBC, we move from Toeplitz matrices to Toeplitz plus Hankel ones,  that is a brand new field of mathematics. 
The formula generalizing FH has been proved very recently \cite{idk-09} (see also \cite{basor2}), 
and there is no conjecture for the subleading terms.
Putting together the ingredients of the generalized FH and the recent results for Toeplitz plus Hankel, we conjecture
a generalized FH  formula for Toeplitz plus Hankel matrices, that we use to determine analytically the corrections to the 
scaling of the entanglement entropy for OBC.

\subsection{Entanglement entropy, conformal field theory, and boundaries}

Let us consider an infinite one-dimensional critical system whose scaling limit is described by a CFT of central charge c, and 
a partition into a finite block of length $\ell$ and the remainder. The entanglement entropy 
(the von Neumann entropy $-\Tr \rho_A \log \rho_A $ of the reduced density matrix $\rho_A$) for 
$\ell$ much larger than the short-distance cutoff $a$ is asymptotically \cite{Holzhey,Vidal,cc-04,cc-rev} 
\be
S_1 \equiv-{\rm Tr} {\rho_A\ln\rho_A}\simeq \frac{c}3 \ln \frac{\ell}a +c'_1\,,
\label{criticalent}
\ee
where  $c'_1$ is a non-universal additive constant.
A more general measure of the quantum entanglement  is provided by
the R\'enyi entropies, that in the conformal situation of before scale like \cite{cc-04,cc-rev}  
\be
S_n\equiv\frac1{1-n}\ln{\rm Tr}\,\rho_{\cal A}^n \simeq \frac{c}6\left(1+\frac1n \right)\ln \ell+c'_n
\, ,
\label{Sndef}
\ee
where $c_n'$ are other non-universal constants.
For $n=1$, Eq. (\ref{Sndef}) reproduces  $S_1$, but  the knowledge of $S_n$ for different $n$ contains more information and 
characterizes the full spectrum of non-zero eigenvalues of $\rho_A$ \cite{cl-08}, that is
fundamental to understand the scaling of some numerical algorithms based on matrix product states \cite{mps}. 

\begin{figure}[t]
\includegraphics[width=\textwidth]{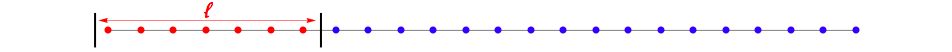}
\includegraphics[width=\textwidth]{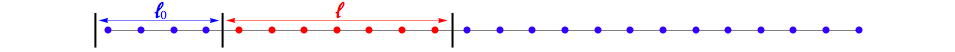}
\caption{The two block configurations we consider.
Top: a block of length $\ell$ starting from the boundary.
Bottom: a block of length $\ell$ at distance $\ell_0$ from the boundary.}
\label{fig:chain}
\end{figure}

For semi-infinite systems, R\'enyi entropies depend on where the block is placed because of the breaking of 
translational invariance. When the block starts from the beginning of the semi-infinite chain (see Fig. \ref{fig:chain}),
very general results are available from boundary CFT. 
The R\'enyi entropies  behave like \cite{cc-04,cc-rev}
\be
S_n \simeq \frac{c}{12}\left(1+\frac1n \right)\ln \ell+\tilde{c}'_n\, ,
\label{snb}
\ee
The constants $\tilde c'_n$ are  non-universal, but their value is related to $c'_n$ for systems with PBC by the {\it universal}
relation  \cite{cc-04,zbfs-06,lsca-06}
\be
{\tilde c}'_n-\frac{c'_n}2= \log g\,,
\label{Bent}
\ee
where $\log g$ is the boundary entropy, first discussed by
Affleck and Ludwig \cite{al-91}. $g$ depends only on the boundary CFT
and its value is known in the simplest cases.
For the XX model with OBC considered here, we have $g=1$ \cite{al-91}.

When the block of length $\ell$ 
placed at a distance $\ell_0$ from the boundary (see Fig. \ref{fig:chain}), the situation is more complicated. 
In this case,  global conformal invariance only gives the general scaling of 
$S_n$ as
\be
S_n= \frac{c}{12}\left(1+\frac1n\right) \log \frac{(2\ell_0 +\ell)^2}{\ell^2 4\ell_0(\ell+\ell_0)} 
+2 \tilde{c}'_n+\frac{\log  {F}_{n}(x)}{1-n} \,,
\label{Fn}
\ee 
where $x$ is the four-point ratio 
\begin{equation}
x=\frac{\ell^2}{(2\ell_0+\ell)^2}\,.
\label{4pR}
\end{equation}
This formula can be readily obtained from the entanglement of two disjoint intervals in an infinite 
system \cite{fps-09,cct-09}, where 
the points corresponding to the second interval are the mirror images (with respect to the boundary) of the 
actual interval.
The function ${F}_{n}(x)$ (normalized such that ${F}_{n}(0)=1$) depends on the full operator content of the theory and 
must be calculated case by case.
The entanglement of two blocks attracted in the last year an enormous interest \cite{fps-09,cct-09,cg-08,atc-09,h-10,c-10} 
because of its ability to detect the full
operator content of a CFT. However, all the results presented so far are for periodic systems  and still little attention has been 
devoted to the boundary case that we address here in a specific case (that we will see to turn out to be trivial, after the
corrections to the scaling have been properly taken into account).

\subsection{Beyond CFT}

When $S_n(\ell)$ is computed numerically, 
it has been observed that the asymptotic CFT result is obscured by large, and often oscillatory, corrections to the scaling
\cite{lsca-06,ccen-10,osc}. 
In Ref. \cite{ccen-10}, on the basis of both exact and numerical results, it has been argued that
these corrections are in fact \emph{universal} and encode information
about the underlying CFT beyond what is captured by the central
charge alone. More precisely, they give access to the scaling dimensions
of some of the most relevant operators of the underlying CFT.  
This conjecture of Ref. \cite{ccen-10} has been recently confirmed
by using perturbed CFT arguments \cite{cc-10} (and generalized to gapped chains as well \cite{ccp-10}). 
For a Luttinger liquid, the proposed scaling form of $S_n$ is \cite{ccen-10,cc-10}
\be
S_n=S_n^{CFT}+ f_n \cos(2k_F \ell) \ell^{-2K/n}\,,
\label{corrPBC}
\ee
where $K$ is the scaling dimension of a relevant operator (in general the oscillating factor can be different from $\cos(2k_F \ell)$ 
or even be absent, as it happens for the Ising model \cite{ce-10,ij-08}).
The constant $f_n$ is a non-universal quantity that has been determined exactly only for XX 
and Ising models \cite{ccen-10,ce-10}.
It is worth mentioning that in the known PBC cases $f_1$ turns out to be zero, i.e. for the entanglement entropy there are no 
unusual corrections. This is also the case for all the numerical computations presented so far, but this fact still 
lacks of a general proof.

On the basis of universality \cite{ccen-10} and directly by CFT \cite{cc-10}, it has been argued that in the case 
of OBC, the exponent governing the corrections is half of the PBC one (i.e. $K/n$ replaces $2K/n$ in Eq. (\ref{corrPBC})), 
as compatible with all numerical computations available \cite{lsca-06,cv-10,cv-10b}. 
It is important to mention that with OBC, unusual corrections are also present for $n\to1$ \cite{lsca-06}. 
In this manuscript, we present a first analytic computation of these corrections to the scaling for systems with OBC. 

We anticipate here the main result of the manuscript for the R\'enyi entanglement entropy of a block starting from the boundary of a 
semi-infinite system
\bea\label{intro}
\fl S_n(\ell)&=&\frac{1}{12}\left(1+\frac1n\right)\ln \left[2 \left(2\ell+1\right)  |\sin k_F|\right] +\frac{E_{n}}{2}\nn\fl
&&+
 \frac{2\sin[k_F (2\ell+1)]}{1-n}
[2(2\ell+1)|\sin k_F|]^{-1/n}\frac{\Gamma(\frac{1}{2}+\frac{1}{2n})}{\Gamma(\frac{1}{2}-\frac{1}{2n})}
+o(\ell^{-1/n})\,,
\eea
that is compatible with the general Luttinger liquid prediction \cite{ccen-10,cc-10} with $K=1$.
Notice that away from half-filling ($k_F=\pi/2$), the  oscillations have  different forms compared to the PBC case.
While this formula is correct only to order $o(\ell^{-1/n})$, in the following we will present the full expansion up to 
the order $o(\ell^{-1})$, for any finite value of $n$. In the case of $n\to\infty$, the corrections become logarithmic and 
are also exactly calculated. 
CFT is also used to infer exact analytic formulas for finite systems with two open boundaries.

\subsection{Organization of the manuscript}

The remainder of this paper is organized as follows. We first present known results for the XX model and the 
calculation of corrections to the scaling with PBC in Sec. \ref{defsec}. 
In Sec. \ref{sec:as} we derive the asymptotic behavior of R\'enyi entropies, 
 while in the Sec. \ref{sec:corr} we calculate analytically the first correction to the 
scaling and a family of the subleading ones that, independently of $n$, give the correct behavior at $o(\ell^{-1})$.
In Sec. \ref{sec:num} we generalize the results to finite systems. 
In Sec. \ref{sec:disc} we consider the case of a block disconnected from the boundary.
Finally in Sec. \ref{sec:concl}, we summarize our main results and discuss problems deserving further investigations.

\section{Entanglement entropy in the XX model}
\label{defsec}
The Hamiltonian of the XX model for a semi-infinite chain  with an open boundary is
\be
H = -  \sum_{l=1}^{\infty} {1\over 2} \left[  \sigma^x_l \sigma^x_{l+1} +  \sigma^y_l \sigma^y_{l+1}\right] - 
h 
\sigma^z_l , 
\label{HXX}
\ee
where $\sigma_l^{x,y,z}$ are the Pauli matrices at site $l$. The
Jordan-Wigner transformation 
\be
c_l=\left(\prod_{m<l} \s^z_m\right)\frac{\s^x_l+i\s_l^y}2\,,
\ee
maps this model to a quadratic Hamiltonian of spinless fermions 
\be
H =
-\sum_{l=1}^{\infty} c_l^\dagger c_{l+1}  + c_{l+1}^\dagger c_{l} + 
2h 
\left(c_l^\dagger c_{l}-\frac{1}{2}\right). 
\label{Hfermi}
\ee
Here $h$ represents the chemical potential for the spinless fermions
$c_l$, which satisfy canonical anti-commutation relations
$\{c_l,c^\dagger_m\}=\delta_{l,m}$. The Hamiltonian (\ref{Hfermi}) is
diagonal in momentum space and for $|h|<1$ the ground-state is a
partially filled Fermi sea with Fermi-momentum 
\be
k_F=\arccos |h|.
\ee
In the following we will always assume that $|h|<1$ so that we are
dealing with a gapless theory.

Using Wick theorem, the reduced density matrix of a block $A= [\ell_0+1,\ell_0+\ell]$ composed of $\ell$ contiguous sites 
in the ground state of the Hamiltonian (\ref{HXX}) can be written as 
\be
\rho_{ A}=\det{C}\
\exp\left(\sum_{j,l\in{ A}}\big[\ln(C^{-1}-1)\big]_{jl}
c^\dagger_jc_l\right),
\ee
where the \emph{correlation matrix} $C$ has matrix elements defined by
\be
C_{nm}= \langle c_m^\dagger c_n \rangle
\label{Cnm}
\ee
In the ground-state of the open  semi-infinite Hamiltonian (\ref{HXX}), the elements of the correlation matrix are \cite{pe-rev} 
\be
C_{nm}=\frac{\sin\bigl(k_F(n-m)\bigr)}{\pi (n-m)}-\frac{\sin\bigl(k_F(n+m)\bigr)}{\pi (n+m)}\,,
\label{Cmno}
\ee
and the PBC result is recovered when $n,m\to \infty$ while keeping their distance $n-m$ finite, i.e. far from the boundary as 
the physical intuition suggests.
The first half of this expression (Toeplitz part) is the same as in systems with periodic boundary conditions, while the second 
part (Hankel type) is direct consequence of the non-translational invariant terms introduced by the boundary.

As a real symmetric matrix, $C$ can be diagonalized by an orthogonal 
transformation
\be
R C R^T\equiv \delta_{lm} (1+\nu_m)/2\,, 
\ee
and the eigenvalues depend both on $\ell$ and $\ell_0$. 
The reduced density matrix $\rho_A$ is uncorrelated in the transformed basis, so that  the R\'enyi entropies can
be expressed in terms of the eigenvalues $\nu_l$ as 
\be\fl
S_n(\ell_0,\ell)=\sum_{l=1}^\ell e_n(\nu_l)\,, \quad {\rm with }\quad
e_n(x)=\frac1{1-n}\ln \left[\left(\frac{1+x}2\right)^n+ \left(\frac{1-x}2\right)^n\right]\,.
\label{Sn}
\ee
More details about this procedure can be found in, e.g., Refs. \cite{Vidal,gl-rev,pe-rev}. 
The above construction refers to the block entanglement of fermionic degrees of freedom. 
However, 
the Jordan-Wigner transformation, although non local,   
mixes only spins inside the block. 
As well known, this ceases to be the case when two or more disjoint intervals are considered \cite{atc-09,ip-09} and other techniques
need to be employed \cite{fc-10} in order to recover CFT predictions \cite{fps-09,cct-09,atc-09}.  
Furthermore, also in the case of XX chains with different boundary conditions (e.g. fixed)  the Jordan-Wigner string 
would spoil the correspondence between spins and fermions for an interval 
detached from the boundary (i.e. $\ell_0\neq0$). 
It is a peculiarity of OBC that the reduced density matrix of any interval at any distance from the 
boundary is the same for spins and fermions.

The representation (\ref{Sn}) is particularly convenient for numerical
computations: the eigenvalues $\nu_m$ of the $\ell\times\ell$
correlation matrix $C$ are determined by standard linear algebra
methods and $S_n(\ell)$ is then computed using Eq. (\ref{Sn}). 
All the numerical computations presented in the following have been obtained in this way.

The sum in Eq. (\ref{Sn}) can be put in the form of an integral on the complex plane \cite{jk-04}, 
introducing the determinant 
\be
D_\ell(\l)=\det\big((\l+1) I -2 C\big)\equiv \det(G)\,.
\label{Ddef}
\ee
In the eigenbasis of $C$ the determinant is simply a polynomial of
degree $\ell$ in $\l$ with zeros $\{\nu_j|j=1,\ldots,\ell\}$, i.e.
\be
D_\ell(\l)=\prod_{j=1}^\ell (\l-\nu_j).
\ee
This implies that the R\'enyi entropies have the integral representation
\be
S_n(\ell)= \frac1{2\pi i}\oint d\l\ e_n(\l) \frac{d\ln D_\ell(\l)}{d\l}\ ,
\label{Snint}
\ee
where the contour of integration encircles the segment $[-1,1]$.
In the PBC case and in the thermodynamic limit ($L\to\infty$), 
Fisher-Hartwig conjecture allows to obtain the asymptotic large $\ell$ behavior of $S_n(\ell)$ \cite{jk-04}.
The generalized Fisher-Hartwig conjecture permits the computation of all harmonic corrections \cite{ccen-10,ce-10}, while 
non-harmonic corrections can be computed only exploiting random matrices techniques \cite{ce-10}.
In next subsection, we report the Fisher-Hartwig approach to PBC, in order to fix the notation and to understand the needed 
ingredients for OBC.

\subsection{The asymptotic result for periodic boundary condition.}

For PBC, the correlation matrix in the limit $L\to\infty$ is
\be
C_{nm}=\frac{\sin\bigl(k_F (n-m)\bigr)}{\pi (n-m)}\,.
\ee 
The matrix $G$ in Eq. (\ref{Ddef})  is a $\ell\times \ell$ Toeplitz matrix, i.e. its
elements depend only on the difference between row and column indices 
$G_{jk}=g_{j-k}$. In this case, it is possible to use the (generalized) Fisher-Hartwig conjecture to calculate the 
asymptotic behavior of $D_{\ell}(\lambda)$ and hence the R\'enyi entropies \cite{jk-04,ce-10}.
The standard FH calculation proceeds as follows. 
We define the \emph{symbol} of the Toeplitz matrix  the
Fourier transform $g(\theta)$ of $g_l$ 
\bea
g_l&=&\int_0^{2\pi}\frac{d\theta}{2\pi}\ e^{il\theta}\ g(\theta),
\label{gll}
\eea
that in our case takes the form
\be
g(\theta)=
\cases{
\lambda+1 & $\theta\in[k_F,2\pi-k_F]$\\
\lambda -1 & $\theta\in [0,k_F]\cup[2\pi-k_F,2\pi]$\ .
}
\label{symbol0}
\ee
On the interval $[0,2\pi]$ the function $g(\th)$ has two
discontinuities at $\th_1=k_F$ and $\th_2=2\pi-k_F$. 
In order to employ the Fisher-Hartwig conjecture 
one needs to express  $g(\theta)$ in the form
\be
g(\th)=f(\th)\prod_{r=1}^R e^{ib_r[\th-\th_r-\pi\sgn(\th-\th_r)]}
\left(2-2\cos(\th-\th_r)\right)^{a_r},
\label{symbol}
\ee
where $R$ is an integer, $a_r$, $b_r$ and $\theta_r$ are constants and
$f(\theta)$ is a smooth function with winding number zero.
The Fisher-Hartwig conjecture then states that the large-$\ell$
asymptotic behavior of the Toeplitz determinant is 
\be
D_\ell\sim F[f(\theta)]^\ell\left(\prod_{j=1}^R \ell^{a_j^2-b_j^2} \right) E\,,
\ee 
where $F[f(\theta)]=\exp(\frac1{2\pi} \int_0^{2\pi} d\th \ln f(\th))$
and $E$ is a known function of 
$f(\th)$, $a_r$, $b_r$, and $\th_r$. In our case it is straightforward
to express the symbol in the canonical form (\ref{symbol}).
As $g(\theta)$ has two discontinuities in $[0,2\pi)$ we have $R=2$.
By comparison, we have
\bea
a_{1,2}&=&0\,,\nonumber \\
b_2&=&-b_1=\beta_\lambda+m\,,\nonumber\\
f(\th)&=&f_0=(\lambda+1)e^{-2ib_2k_F}=(\lambda+1)e^{-2ik_F m}e^{-2ik_F\beta_\l}\,,
\label{FHc}
\eea
where $m$ is an arbitrary integer number, that labels the different inequivalent
\emph{representations} of the symbol $g(\th)$, see \cite{bm-94}, and 
\be
\beta=\frac{1}{2\pi i}\log \Bigl[\frac{\lambda+1}{\lambda-1}\Bigr]\qquad -\pi\leq \arg{\frac{\lambda+1}{\lambda-1}}<\pi\, .
\label{betadef}
\ee 
Jin and Korepin employed the Fisher-Hartwig conjecture
for the $m=0$ representation and obtained the following result for the
large-$\ell$ asymptotics of $D_\ell(\lambda)$ \cite{jk-04}
\be\fl
D_\ell^{JK}(\l)\sim
\left[(\l+1)\left(\frac{\l+1}{\l-1}\right)^{-\frac{k_F}\pi}\right]^\ell \
(2 \ell |\sin k_F|)^{-2\beta^2(\l)} G^2(1+\beta_\l)G^2(1-\beta_\l)\,,
\label{Djk}
\ee
where $G(x)$ is the Barnes G-function \cite{BG}.
Inserting (\ref{Djk}) into (\ref{Snint}) and carrying out the integral
leads to the result for the asymptotic behaviour of the
R\'enyi entropy 
\be
S_n^{JK}(\ell)=\frac16\left( 1+\frac1n\right)\ln (2\ell |\sin
k_F|)+E_n\,, 
\label{SnJK}
\ee
where the constant $E_n$ has the integral representation
\be\fl
E_n=\left(1+\frac1n\right)\int_0^\infty \frac{dt}t 
\left[\frac{1}{1-n^{-2}}
\left(\frac1{n\sinh t/n}-\frac1{\sinh t}\right)
\frac1{\sinh t}-\frac{e^{-2t}}6\right]\, .
\label{cnp}
\ee

However, when the symbol has several inequivalent representations as in our case, the 
{\it generalized} Fisher-Hartwig conjecture 
(gFHC) \cite{bm-94} applies and one has to sum over all these representations as 
\be
D_\ell(\l)\sim\sum_m (f_0(m))^{\ell}
\ell^{-\sum_{r=1}^2(b_r(m))^2}E(m)\,,
\label{Drep}
\ee
where all the various FH constants $f_0$, $b_i$, and $E$ depend on $m$, as shown in Eq. (\ref{FHc}).
Then the full result of the generalized Fisher-Hartwig conjecture for the
Toeplitz determinant takes the form \cite{ce-10}
\bea\fl
D_\ell&\sim&(\l+1)^\ell\left(\frac{\l+1}{\l-1}\right)^{-\frac{k_F\ell}\pi}
\sum_{m\in\mathbb{Z}}
(2\ell |\sin k_F|)^{-2(m+\beta_\l)^2} e^{-2ik_Fm\ell}\nn
\fl&&\hskip 4cm\times \left[G(m+1+\beta_\l)G(1-m-\beta_\l)\right]^2,
\label{Drep2}
\eea
leading after long calculations for the integral (\ref{Snint}) to \cite{ce-10}
\be
\fl S_n(\ell)-S^{JK}_n(\ell)=\frac{ 2 \cos(2 k_F \ell) }{1-n} (2\ell
|\sin k_F |)^{-2/n} 
\left[\frac{\Gamma(\frac{1}{2}+\frac{1}{2n})}
{\Gamma(\frac{1}{2}-\frac{1}{2n})}\right]^2
+o\big(\ell^{-2/n}\big).
\label{Snfinal}
\ee
We mention that the gFH conjecture has also other applications in physics as for example those reported in Refs. \cite{ov,fa-05}.

\section{Fisher-Hartwig like conjecture for the semi-infinite chain}
\label{sec:as}


We want to use a generalization of the FH conjecture to obtain the determinant in Eq. (\ref{Ddef}) for large $\ell$. 
The matrix $G$ is  not of the Toeplitz form because of the second term in $C_{nm}$ in Eq. (\ref{Cmno}). 
Let us consider a block at distance $\ell_0$ from the boundary (i.e. starting from the site $\ell_0+1$). 
To fix the notation,  the matrix $G$ is an $(\ell+1)\times (\ell+1)$ matrix that is a sum of 
Toeplitz and Hankel matrices with elements of the form 
\be
G_{nm}= g_{n-m}- g_{n+m+2\ell_0+2} \,,\qquad n,m=0,1,\dots \ell-1 \,.
\label{Gnmo}
\ee
Here $g_l$ is the same as in Eq. (\ref{gll}).
We are not aware of any possible generalization (conjectured or proved) of the FH formula to
the case of arbitrary $\ell_0$. 
However,  very recently a theorem has been 
proved \cite{idk-09} for some Toeplitz+Hankel matrices, among which those of  the form 
\be
a_{i-j}-a_{i+j+2}\,, 
\ee
that applies to our case with $\ell_0=0$, i.e. when the block starts from the boundary.

For the case of a piecewise constant symbol as the one of our interest (cf. Eq. (\ref{symbol0})), this theorem was known 
from longer time \cite{basor2}. 
The result in Refs. \cite{idk-09,basor2} has a structure similar to the FH formula, with more constants $a_j,b_j$ corresponding to boundary terms. 
Such a general formula is not very illuminating and too long to be written in its full glory. 
We remand the interested reader to the original reference \cite{idk-09}. 
We specialize this formula to the symbol in Eq. (\ref{symbol0}), and we obtain the asymptotic behavior of 
the determinant $D_{\ell}(\lambda)$ as
\be\fl
D_\ell(\l)\sim e^{i(\frac{\pi}{2}-k_F)\beta} \Bigl[(\lambda+1)\Bigl(\frac{\lambda+1}{\lambda-1}\Bigr)^{-\frac{k_F}{\pi}}\Bigr]^{\ell}(4\ell|\sin(k_F)|)^{-\beta^2}G(1-\beta)G(1+\beta)\,,
\ee
where $\beta$ has been defined in Eq. (\ref{betadef}).
The result is very similar to the PBC case. The main differences are in the halving of the exponent of $\ell$,
in the factor $4$ instead of $2$ multiplying $\ell$ (both with a clear physical interpretation), 
in the absence of the square in the Barnes G function,  and in the phase shift in front.

We can now proceed to calculate the integral (\ref{Snint}) giving the asymptotic R\'enyi entropies. 
First we notice that the phase shift
$e^{i(\frac{\pi}{2}-k_F)\beta}$
in Eq. (\ref{eq:Dlleading}) gives a \emph{vanishing} contribution to the entropies, indeed
\be
\Bigl(\frac{1}{2}-\frac{k_F}{\pi}\Bigr)\frac{1}{2\pi i}\oint\mathrm d \lambda \frac{e_n(\lambda)}{1-\lambda^2}=0\, .
\ee
The remaining part of the integral parallels that for PBC, giving 
\be
S_n(\ell)=\frac{1}{12}\left(1+\frac1n\right)\ln (4 \ell |\sin k_F|) +\frac{E_{n}}{2} \, ,
\label{Snobcl}
\ee
where $E_n$ is defined in Eq. (\ref{cnp}). 

This result  for $S_n$ is exactly what expected from CFT, as reviewed in the introduction. 
Both the leading logarithmic term and the subleading constant one are in agreement with Eq. (\ref{snb}) with $\ln g=0$. 
There is then no new physical information in this expression.
However, the present result is based on a mathematical  theorem and so it provides a rigorous confirmation of a general CFT result
in a specific lattice model.
The new physical results of this paper are  in the following sections.

\section{Generalized conjecture for the oscillating corrections  to the scaling}
\label{sec:corr}

We consider now the corrections to the scaling to the asymptotic result derived in the previous section. 
Analogously to the PBC case, we need a generalized Fisher-Hartwig formula that applies to the case
of Toeplitz+Hankel matrices. 
To the best of our knowledge, this formula does not exist neither conjectured or proved. 
However, armed with the general ideas presented above is not difficult to give our own conjecture for such a determinant: 
we should only sum on the several different inequivalent representations of the symbol.  

\begin{figure}[t]
\includegraphics[width=0.55\textwidth]{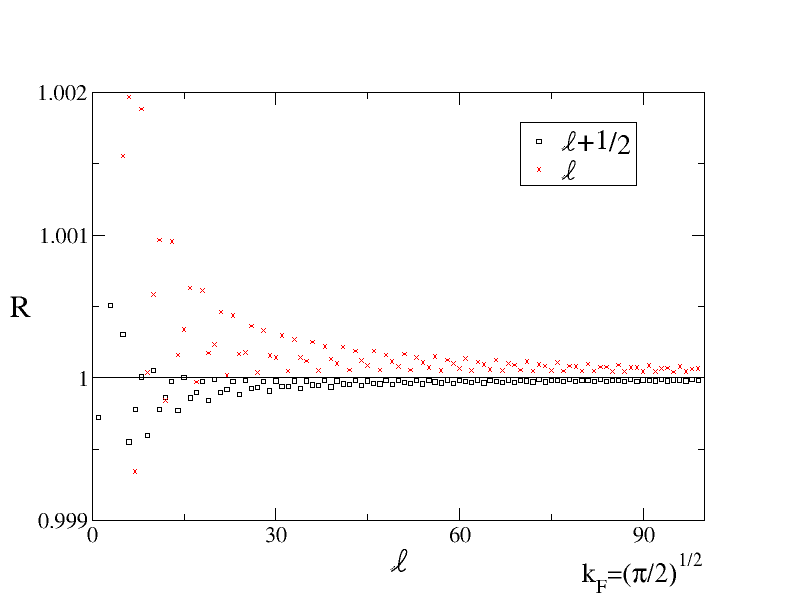}
\includegraphics[width=0.55\textwidth]{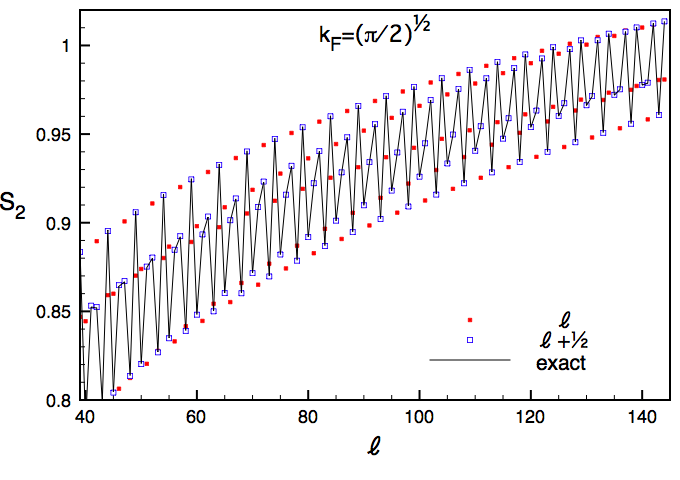}
\caption{Left: The ratio $R=D_{\ell}(\lambda)/D^{(0)}_{\ell}(\lambda)$ for $\l=1+i$ and $k_F=\sqrt{\pi/2}$.  
Right: The entanglement entropy for $k_F=\sqrt{\pi/2}$. 
Both graphs show the importance of the additive factor $1/2$.
}
\label{fig:D}
\end{figure}

For the symbol in Eq. (\ref{symbol0}), we {\it conjecture} the following formula
\bea\fl
D_\ell(\l)&\sim&  (\l+1)^\ell\left(\frac{\l+1}{\l-1}\right)^{-\frac{k_F\ell}\pi}
\sum_{m\in\mathbb{Z}}e^{i\frac{\pi}{2}(\beta+m)}
\left[ 4\Bigl(\ell+\frac12\Bigr) |\sin k_F|\right]^{-(m+\beta_\l)^2}  \nn
\fl&&\hskip 3cm\times e^{-2ik_F(\beta+m)(\ell+{1}/{2})} G(m+1+\beta_\l)G(1-m-\beta_\l).
\label{Drep2o}
\eea
Most of the above formula is inspired to the gFH Eq. (\ref{Drep2}) and adapted to the present case. 
However, the factor $1/2$ as an additive constant to $\ell$ has been introduced without any 
mathematical reason. 
This factor $1/2$ gives an analytic (i.e. non-harmonic) correction to $D_{\ell}(\l)$ and we introduced it to reproduce 
accurately the numerical data.
In Fig. \ref{fig:D} we report the ratio of the numerically calculated $D_{\ell}(\l)$ asymptotic value with and without the additional $1/2$, 
showing that the former converges faster. 
The data for the resulting entanglement entropy $S_2(\ell)$ (reported in the right panel of Fig. \ref{fig:D}) show even more clearly 
the importance of this factor.   
In any case, we conjectured this gFH formula and 
in doing so we prefer to conjecture an expression that reproduces numerical data as accurately as possible. 
A full justification of the factor $1/2$ could be obtained by generalizing the random matrix 
results for PBC in Ref. \cite{fw-05}, but this is beyond our knowledge.

The leading corrections to the scaling is obtained by summing only the three modes $m=-1,0,1$,  obtaining
\bea\fl
D_\ell&\sim& D_\ell^{(0)}\Bigl\{1+ie^{-ik_F}e^{-2ik_F \ell}L_k^{-1-2\beta}\frac{\Gamma(1+\beta)}{\Gamma(-\beta)}-ie^{ik_F} e^{2i k_F \ell}L_k^{-1+2\beta}\frac{\Gamma(1-\beta)}{\Gamma(\beta)}\Bigr\}\nonumber\\
\fl&\equiv& D_\ell^{(0)} (1+\Psi_\ell(\l))\,,
\label{Dasy}
\eea
where we isolated the leading term
\be\label{eq:Dlleading}
\fl D_\ell^{(0)}\equiv e^{i(\frac{\pi}{2}-k_F)\beta} \Bigl[(\lambda+1)\Bigl(\frac{\lambda+1}{\lambda-1}\Bigr)^{-\frac{k_F}{\pi}}\Bigr]^{\ell}L_k^{-\beta^2}G(1-\beta)G(1+\beta)\,,
\ee
we defined $\Psi_{\ell}(\l)$, and $L_k$ for OBC is  defined as
\be
L_k=2(2\ell +1)|\sin(k_F)|\, .
\ee
We define 
\be
 d_n(\ell)\equiv S_n(\ell)-S_n^{{0}}(\ell)\,,
\ee
with 
\be
S_n^{(0)}=\frac{1}{12}\left(1+\frac1n\right)\ln \left[2 \left(2\ell+1\right)  |\sin k_F|\right] +\frac{E_{n}}{2} \,,
\ee
that includes the additive $1/2$ factor compared to the leading term in Eq. (\ref{Snobcl}). 
For large $L_k$ we have 
\bea
\fl d_n(\ell)&\sim& \frac1{2\pi i}\oint d\lambda\ e_n(\l) \frac{d\ln 
\left[1+ \Psi_\ell(\lambda)\right]}{d\l} 
=\frac1{2\pi i}\oint d\lambda\ e_n(\l) \frac{d
\Psi_\ell(\lambda)}{d\l}+\ldots.
\label{approx}
\eea
The contour integral can be written as the sum of two contributions
infinitesimally above and below the interval $[-1,1]$ respectively, i.e. 
\bea
d_n(\ell)&\sim&
\frac1{2\pi i}\left[\int_{-1+i\epsilon}^{1+i\epsilon}-
  \int_{-1-i\epsilon}^{1-i\epsilon}\right]d\lambda\
 e_n(\l) 
 \frac{d\Psi_\ell(\lambda)}{d\l} .
\label{int2}
\eea
This shows that we only require the discontinuity across the branch
cut. The only discontinuous function is $\beta_\l$, which for 
$-1<x<1$ behaves as
\be
\beta_{x\pm i\epsilon}= -i w(x) \mp \frac12\,, \qquad {\rm with}\qquad 
w(x)=\frac1{2\pi} \ln \frac{1+x}{1-x}\,.
\ee
We now change variables from $\lambda$ to $w$
\be
\l= \tanh(\pi w)\ ,\quad -\infty<w<\infty.
\ee
We have
\bea\fl
\left[L_k^{-1-2\beta} \frac{\Gamma(1+\beta)}{\Gamma(-\beta)}\right]_{\beta=-i w-\frac12}-
\left[L_k^{-1-2\beta} \frac{\Gamma(1+\beta)}{\Gamma(-\beta)}\right]_{\beta=-i w+\frac12}&\simeq&
L_k^{2 i w} \gamma(w),
\nn \fl
\left[L_k^{-1+2\beta} \frac{\Gamma(1-\beta)}{\Gamma(\beta)}\right]_{\beta=-iw-\frac12}-
\left[L_k^{-1+2\beta} \frac{\Gamma(1-\beta)}{\Gamma(\beta)}\right]_{\beta=-iw +\frac12}&\simeq&
-L_k^{-2 i w} \gamma(-w), 
\nonumber
\eea
where we have dropped terms of order $O(L_k^{-2})$ compared to the
leading ones and we have defined
\be
\gamma(w)=\frac{\Gamma(\frac12-i w)}{\Gamma(\frac12+i w)}.
\ee
Integrating by parts and using
\be
\frac{d}{dw}e_n(\tanh(\pi w))=
\frac{\pi n}{1-n}(\tanh(n\pi w)-\tanh(\pi w))\,,
\label{intfin}
\ee
we arrive at
\bea\fl
d_n(\ell)&\sim&\frac{i n}{2(1-n)}\int_{-\infty}^{\infty}\mathrm{d}w (\tanh(\pi w)-\tanh(n\pi w))
\nonumber\\ \fl &&\hspace{1cm}
\Bigl[ie^{-ik_F}e^{-2i k_F \ell}L_k^{2i w}\frac{\Gamma(\frac{1}{2}-iw)}{\Gamma(\frac{1}{2}+i w)}+ie^{i k_F}e^{2i k_F \ell}L_k^{-2i w}\frac{\Gamma(\frac{1}{2}+i w)}{\Gamma(\frac{1}{2}-i w)}\Bigr]\, ,
\eea
For large $\ell$ the leading contribution to the integral arises from
the poles closest to the real axis. These are located at
$w_0=i/2n$ ($w_0=-i/2n$) for the first (second) term.
Evaluating their contributions to the integral gives
\be\label{eq:osccorr}
 d_n(\ell)\sim \frac{2\sin [k_F (2\ell+1)]}{1-n}
\Bigl[2(2\ell+1)|\sin k_F|\Bigr]^{-1/n}\frac{\Gamma(\frac{1}{2}+\frac{1}{2n})}{\Gamma(\frac{1}{2}-\frac{1}{2n})}\, ,
\ee

\begin{figure}[t]
\includegraphics[width=0.55\textwidth]{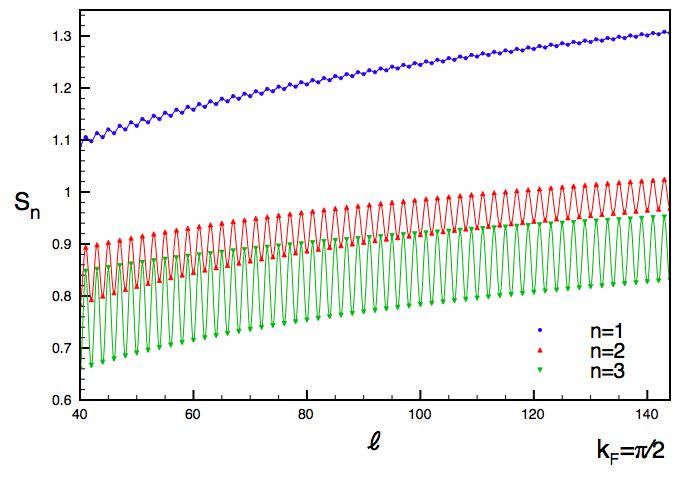}
\includegraphics[width=0.55\textwidth]{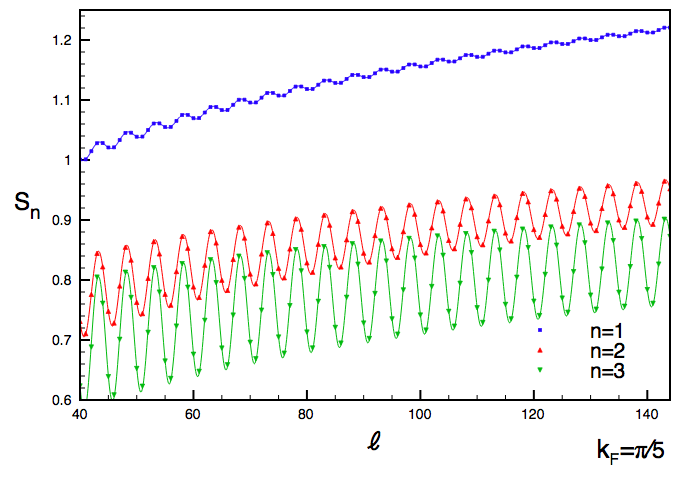}
\includegraphics[width=0.55\textwidth]{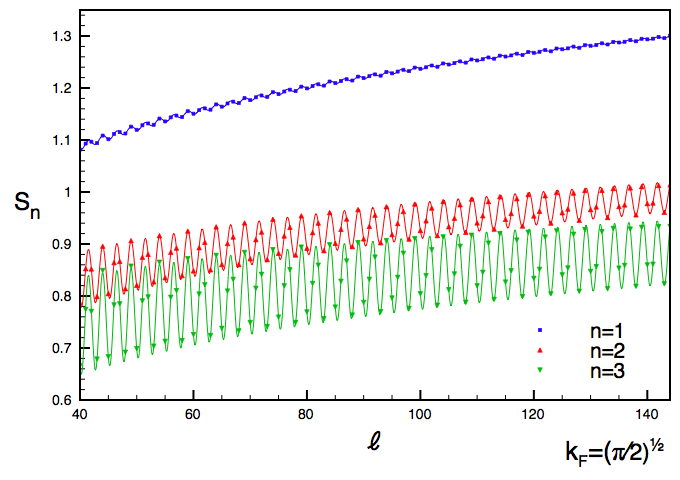}
\includegraphics[width=0.55\textwidth]{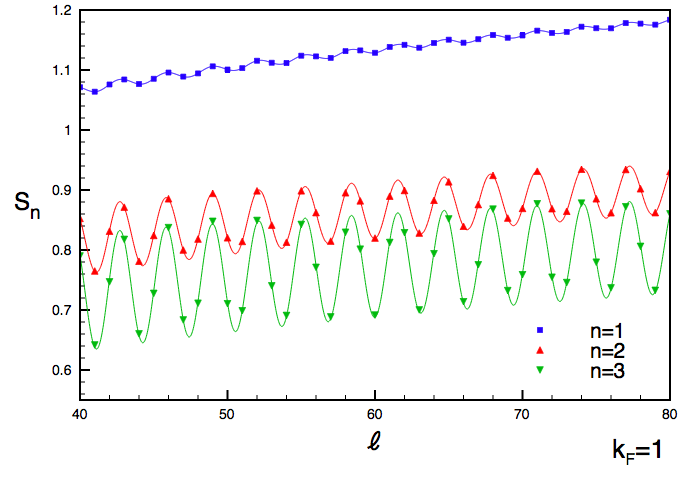}
\caption{$S_n(\ell)$ in a semi-infinite XX chain for $n=1,2,3$ and different values of $k_F$ specified in the captions.
The exact numerical results are compared with the asymptotic formula including only the 
first correction to the scaling. For these small values of $n$, the agreement is excellent.  
Notice the incommensurability effects when $k_F$ is not a fraction of $\pi$ (lower graphs). 
}
\label{fig:TD}
\end{figure}

In Fig. \ref{fig:TD} we report the numerical calculated $S_n(\ell)$ for a semi-infinite chain and for small values of $n=1,2,3$.
For these values of $n$, the inclusion of only the first correction to the scaling (as in Eq. (\ref{intro})) is enough to describe very 
accurately $S_n(\ell)$ even for relatively small values of $\ell$. 
The figure shows the correctness also of the $k_F$ dependence of the 
correction, that is the most important difference compared to PBC.
Notice that without the factor ${1}/{2}$ introduced by hand in Eq. (\ref{Drep2o}), 
the correction (\ref{eq:osccorr}) would be inadequate at order $1/\ell$, as anticipated in Fig. \ref{fig:D}.

\subsection{Subleading corrections}

Eq. (\ref{Snfinal}) describes the asymptotic behavior in the limit
$L_k\to\infty$ with $n$ fixed. It provides a good approximation for
large, finite $\ell$ as long as $\ln( L_k)\gg n$. 
For practical purposes it is useful to know the corrections to $S_n(\ell)$ for large $\ell$ but $\ln( L_k)$
not necessarily much larger than $n$. 
In this regime there are two main sources of corrections to (\ref{Snfinal}). 
The integral (\ref{intfin}) is no longer dominated by the poles
closest to the real axis and contributions from further poles need to
be included. 
These give rise to corrections proportional to $L_k^{- q /n}$, with $q$ integer. 
Furthermore, terms in the expansion of the logarithm in
Eq. (\ref{approx}) need to be taken into account. The corresponding
contributions are proportional to $e^{\pm i 2p k_F \ell}$ with
$p=2,3,\dots$.

We now take both types of corrections into account. 
The following derivation is very similar to the one for PBC in Ref. \cite{ce-10}.
We first consider
the series expansion of the logarithm in Eq. (\ref{approx}).
\be
\ln\big[1+\Psi_\ell(\lambda)\big]=\sum_{p=1}^\infty 
\frac{(-1)^{p+1} \big(\Psi_\ell(\lambda)\big)^p}{p}\,.
\ee
Recalling the explicit expression (\ref{Dasy}) for
$\Psi_\ell(\lambda)$ leads to a binomial sum
\bea
\fl\big(\Psi_\ell(\lambda)\big)^p
&=&\left(i e^{ik_F} e^{-2i k_F\ell}L_k^{-(1+2\beta_\l)}c_{\beta_\l} - 
i e^{-ik_F} e^{2i k_F\ell}L_k^{-(1-2\beta_\l)}c_{-\beta_\l}\right)^p\nn 
\fl&=&
\sum_{q=0}^p {p \choose q} i^p (-1)^q e^{i k_F (2q-p)}
e^{2ik_F \ell (2q-p)} L_k^{-p} L_k^{-2(p-2q)\beta_\l}c_{\beta_\l}^{p-q}
c_{-\beta_\l}^{q}\,,
\eea
where we have introduced the shorthand notation
$c_{\beta}=\Gamma(1+\beta)/\Gamma(-\beta)$.   

When calculating the
discontinuity across the branch cut running from $\lambda=-1$ to
$\lambda=1$ all terms other than $q=0$ and $q=p$ give rise to terms
that are subleading in $L_k$. Hence we may approximate 
\bea
\fl
\big(\Psi_\ell&&(\tanh(\pi w)+i\epsilon)\big)^p-
\big(\Psi_\ell(\tanh(\pi w)-i\epsilon)\big)^p \approx\nn 
\fl&& i^p e^{-2ik_F (\ell+1/2) p}L_k^{2iwp}c_{-iw-1/2}^p
-(-i)^p e^{2ik_F (\ell+1/2) p}L_k^{-2iwp}c_{+iw-1/2}^p\,.
\eea
Plugging this into Eq. (\ref{approx}) 
\bea\fl
d_n(\ell)&\sim&\sum_{p=1}^\infty\frac{(-1)^{p+1}}{p}
\frac{i n}{2(1-n)}\int_{-\infty}^\infty dw 
(\tanh(\pi w)-\tanh(n\pi w))\nn\fl  &\times&i^p
\!\left[e^{-2ipk_F (\ell+1/2)}  L_k^{2 i w p} \gamma^{p}(w)
+(-1)^{p+1} e^{2ipk_F (\ell+1/2)}  L_k^{-2 i wp} \gamma^{p}(-w)
\right].
\label{finint2}
\eea
The integral is carried out by contour integration, taking the two
terms in square brackets into account separately. The first (second)
contribution has simple poles in the upper (lower) half plane at
$w_q= i\frac{2q-1}{2n}$ ($w_q= -i\frac{2q-1}{2n}$), where $q$ is a
positive integer such that $2q-1\neq n,3n,5n,\ldots$ for $n\neq1$.
Contour integration then gives
\bea
\fl
d_n(\ell)&=&\frac{2}{1-n}{\sum_{p,q=1}^\infty}
\frac{(-1)^{p+1}}{p} \cos\Big( \frac{p(\pi-2k_F)}2-  2 p k_F \ell \Big)
L_k^{-\frac{p(2q-1)}{n}}
\big(Q_{n,q}\big)^{p}
+\dots\,,
\label{main}
\eea
where we defined the constants $Q_{n,q}$ as
\be
Q_{n,q}=\frac{\Gamma(\frac{1}{2}+\frac{2q-1}{2n})}
{\Gamma(\frac12-\frac{2q-1}{2n})}\,.
\label{Qn}
\ee
In the sum over $q$ the special values $2q-1\neq n,3n,5n,\ldots$ are
to be omitted  for $n\neq1$. 
Eq. (\ref{main}) shows that there are contributions to the R\'enyi entropies with oscillation
frequencies that are arbitrary multiples of $2k_F$.

At half-filling ($k_F=\pi/2$) certain simplifications occur. 
For even $\ell$ we find
\be\fl
d_n(\ell)\sim
\frac{2}{1-n}\left[
(4\ell)^{-\frac{1}{n}}Q_{n,1} -(4\ell)^{-\frac{2}{n}}
\frac{Q_{n,1}^{2}}{2}
 +(4\ell)^{-\frac{3}{n}}\left(\frac{Q_{n,1}^3}{3} +Q_{n,3}\right)
\right]+\dots\,,
\ee
while for odd $\ell$ we obtain
\be\fl
d_n(\ell)\sim
\frac{-2}{1-n}\left[ (4\ell)^{-\frac{1}{n}} Q_{n,1}
+(4\ell)^{-\frac{2}{n}} \frac{Q_{n,1}^{2}}{2}
+(4\ell)^{-\frac{3}{n}}\left(\frac{Q_{n,1}^{3}}{3} +Q_{n,3} \right)
\right]+\dots\,,
\ee
that are of the same form as for PBC \cite{ce-10} with exponents that are halved.
However, we stress that this is true only at half filling, while for generic $k_F$, Eq. (\ref{finint2}) shows 
 oscillations different from its PBC counterpart.

In all the above analysis we have ignored contributions to the
generalized Fisher-Hartwig conjecture with $|m|>1$. While these lead
to oscillatory contributions with frequencies that are integer
multiples of $2k_F$ they are suppressed by additional powers of
$\ell^{-1}$ and hence are subeading, even in the case where $n$ is not
small. 

\begin{figure}[t]
\includegraphics[width=0.55\textwidth]{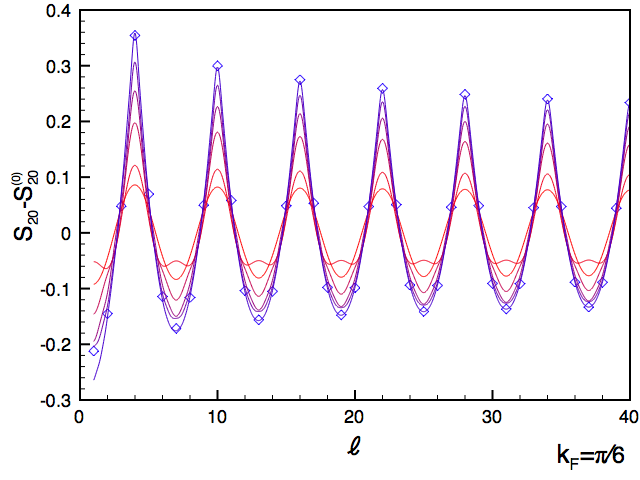}
\includegraphics[width=0.55\textwidth]{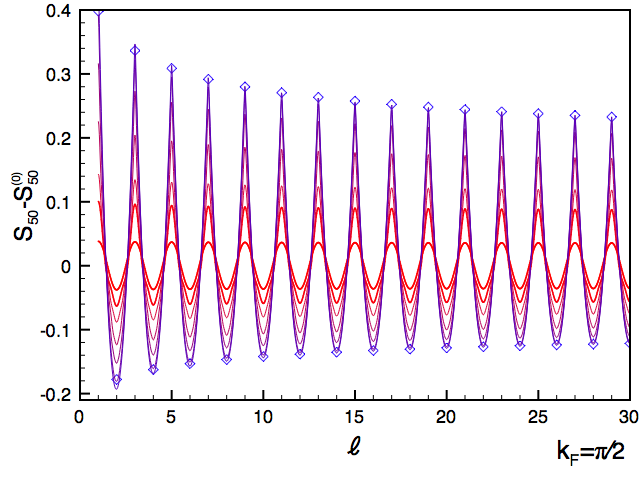}
\caption{Difference of the R\'enyi entanglement entropy with its asymptotic value for $n=20$ (left) and $50$ (right)  
with $k_F = \pi/6$ and $\pi/2$.
The exact numerical data are compared with Eq. (\ref{main}) with increasing number of terms. 
For $n=50$ we report curves with 1,3,5,9,15,25,41 terms, while for $n=20$ we consider 1,2,3,5,8,13 terms.
Increasing the number of terms considered, the expansion (\ref{main}) becomes accurate even for values of $\ell$
as small as 2.}
\label{fig:succ}
\end{figure}

In Fig. \ref{fig:succ} we show  the corrections $d_n(\ell)$ for $n = 20$ and $50$ with $k_F = \pi/6$ and $\pi/2$ 
respectively and their comparison with the asymptotic result Eq. (\ref{main}). 
Step by step we take into account further terms in the asymptotic expression until we obtain a satisfying agreement with the 
numerical data. 
For $n = 20$, 13 terms in Eq. (\ref{main}) are enough to reproduce the data, 
while for $n = 50$ we need 41 terms to have the same accuracy. 
These numbers are larger than the corresponding ones for PBC \cite{ce-10} because the corrections in the present case 
 have smaller exponents.

\subsection{The limit of large $n$}

It is apparent from (\ref{main}) that the limit $n\to\infty$ deserves
special attention. $S_\infty(\ell)$ is known in the literature as {\it single copy entanglement} \cite{sce}.   
Here it is necessary to sum up an infinite number
of contributions in order to extract the large-$\ell$ asymptotics.
It also provides information on the behavior of $S_n(\ell)$ in the regime $n\gg \ln L_k$, $L_k\gg 1$. 
The following derivation parallels the one for PBC in Ref. \cite{ce-10}.

In order to investigate the limit $n\to\infty$ we consider
Eq. (\ref{finint2}), but now first take the parameter $n$ to infinity
and then carry out the resulting integrals. This gives
\bea\fl
d_\infty(\ell)&\sim&
\frac{i}2\sum_{p=1}^\infty \frac{(-1)^p}{p}
\int_{-\infty}^\infty dw (\sgn(w)-\tanh(\pi w))\nn 
\fl&&\times i^p
\left[e^{-2ik_F (\ell+1/2) p} 
L_k^{2 i wp} \left[\gamma(w)\right]^{p}+
(-1)^p e^{2ik_F (\ell+1/2) p} 
L_k^{-2 i wp}\left[
\gamma(-w)\right]^{p}
\right]\nn\fl
&=&-\sum_{p=1}^\infty \frac{(-i)^p}{p}\left[
e^{-2ik_F (\ell+1/2) p}  {\rm Im} \int_{0}^\infty dw 
[1-\tanh(\pi w)]
L_k^{2 i wp} \left[\gamma(w)\right]^{p}
\right. \nn\fl  
&&\left. \qquad\quad-(-1)^p
e^{2ik_F (\ell+1/2) p}  {\rm Im} \int_{0}^\infty dw [1-\tanh(\pi w)]
L_k^{-2 i wp} \left[
\gamma(-w)\right]^{p}
\right].
\eea
Using that the first singularity in the upper (lower) half plane
occurs at $w=i/2$ ($w=-i/2$) we deform the contours to run parallel to
the real axis with imaginary parts $i/4$ and $-i/4$ respectively,
i.e. for the first term we use
$$
\int_0^\infty dw\ f(w)= \int_0^{i/4} dw\ f(w) +\int_{i/4}^{\infty+i/4}
dw\ f(w).
$$
It is straightforward to show that the second integral contributes
only to order $O(1/L_k)$ and does not give rise to logarithmic corrections.
Hence the leading contribution is of the form
\bea\fl
d_\infty(\ell)&\sim&\sum_{p=1}^\infty \frac{(-i)^p}{p}
\left[e^{-2ik_F (\ell+1/2) p}  {\rm Re} \int_{0}^{1/4} dz
(1-i\tan(\pi z))
L_k^{-2 z p} \left(\gamma(iz)\right)^{p}
\right.\nn\fl &&\qquad \left. + (-1)^p
e^{2ik_F (\ell+1/2) p} {\rm Re} \int_{0}^{1/4} dz(1+i\tan(\pi z))
L_k^{-2 z p} \left(\gamma(iz)\right)^{p}
\right]\nn\fl &=&
2\sum_{p=1}^\infty \frac{(-1)^p}{p} \cos\Big( \frac{p(\pi-2k_F)}2-  2 p k_F \ell \Big)
\int_{0}^{1/4} dz e^{-2 z p \ln L_k} 
\left(\gamma(iz)\right)^{p}\ .
\eea
For large $L_k$ the dominant contribution to the integral is obtained by
expanding $\left(\gamma(iz)\right)^{p}$ in a power series around $z=0$
\bea\fl
d_\infty(\ell)&\sim&
2\sum_{p=1}^\infty \frac{(-1)^p}{p} \cos\Big( \frac{p(\pi-2k_F)}2-  2 p k_F \ell \Big)
\int_{0}^{1/4} dz e^{-2 z p \ln L_k} (1+ 2 p z \Psi(1/2)+\dots)\nn\fl
&=&
2\sum_{p=1}^\infty \frac{(-1)^p}{p^2} \cos\Big( \frac{p(\pi-2k_F)}2-  2 p k_F \ell \Big)
\left[\frac1{2\ln L_k}+ \frac{ \Psi(1/2)}{2\ln^2 L_k}+\dots\right],
\label{expansion}
\eea
where $\Psi(z)$ is the digamma function $\Psi(z)=\Gamma'(z)/\Gamma(z)$.

\begin{figure}[t]
\includegraphics[width=0.7\textwidth]{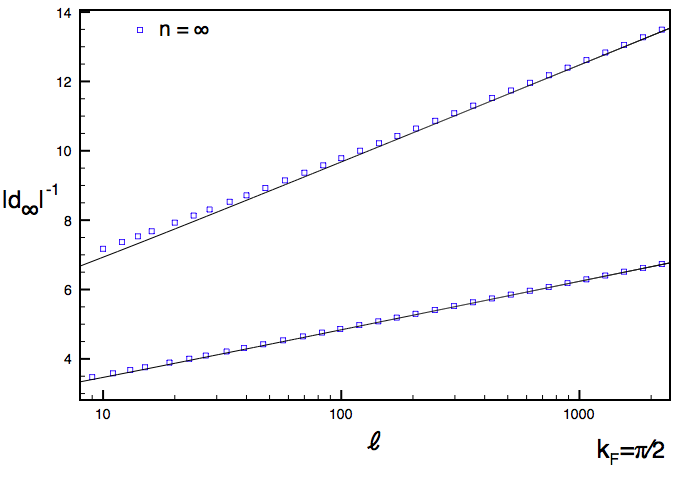}
\caption{$1/|d_\infty(\ell)|$ with  $d_\infty(\ell)=S_\infty(\ell)-S^{(0)}_\infty(\ell)$ vs $\ell$ in log-linear scale for $k_F=\pi/2$. 
Corrections to the scaling are logarithmic and perfectly described by Eq. (\ref{Sinfty}) that is represented as two straight lines 
for $\ell$ even and odd respectively. 
}
\label{fig:Sinf}
\end{figure}

As in the PBC case,  at half-filling  ($k_F=\pi/2$) this result takes
a particularly simple form
\be
d_\infty(\ell)\sim
\frac1{\ln L_k}\sum_{p=1}^\infty \frac{(-1)^{p(\ell+1)}}{p^2}=
\cases{\displaystyle
\frac1{\ln L_k} \frac{\pi^2}6 & $\ell$ {\rm odd} \,,\\ \displaystyle
-\frac1{\ln L_k} \frac{\pi^2}{12} & $\ell$ {\rm even}\,.
}
\label{logcorr}
\ee
Summing some of the subleading terms in (\ref{expansion}) to all
orders in $(\ln(L_k))^{-1}$ leads to an expression of the form
\be
d_\infty(\ell)\sim
\frac{\pi^2}{12\ln(b L_k)}
\cases{\displaystyle
2 & $\ell$ {\rm odd} \,,\\ \displaystyle
-1 & $\ell$ {\rm even}\,,
}
\label{Sinfty}
\ee
where $b=\exp(-\Psi(1/2))\approx 7.12429$. 
To check this result, in Fig. \ref{fig:Sinf}, we report the exact numerical data (only for $k_F=\pi/2$) for $1/|d_\infty(\ell)|$ in 
log-linear scale, showing explicitly the logarithmic form of the corrections, described very precisely by Eq. (\ref{Sinfty}).

\section{Finite systems}
\label{sec:num}

The Hamiltonian of a finite XX chain with two open boundaries is
\be
H = -  \sum_{l=1}^{L-1} {1\over 2} \left[  \sigma^x_l \sigma^x_{l+1} +  \sigma^y_l \sigma^y_{l+1}\right] - 
h \sum_{l=1}^L  \sigma^z_l \,. 
\label{HXXf}
\ee
As before, the Hamiltonian is diagonalized by a Jordan-Wigner transformation and Fourier transform. 
However, we are dealing with a finite chain at fixed magnetic field, that in the language of fermions is fixed chemical potential. 
In this case, the number of fermions is a non-continuous function of $L$, 
because it can assume only integer values.
The number of fermions is $N_F= \lfloor (L+1) |\arccos h| \rfloor$. 
There is an ambiguity in defining the Fermi momentum and we choose the definition 
\be
k_F\equiv \frac{\pi N_F}{L}= \frac{\pi}L \lfloor (L+1) |\arccos h| \rfloor\,,
\ee
i.e. $k_F$ is an integer multiple of $\pi/L$.
This definition has the advantage to give $k_F=\pi/2$ at half-filling for $L$ even. 
However, the correlation matrix $C$ is better defined in terms of 
\be
k'_F= k_F \frac{L}{L+1}+\frac{\pi}{2(L+1)}\,.
\ee
Notice the following important properties: 
1) In the limit $L\to\infty$, $k_F=k_F'$; 
2) If $k_F=\pi/2$, then $k'_F=\pi/2$;
3) For $L$ odd, the ground state is doubly degenerate and both $k_F$ and $k'_F$ cannot be equal to $\pi/2$.   
With this definition, we can write the correlation matrix as
\be
C_{nm}=\frac{1}{2(L+1)}\left[\frac{\sin\bigl(k'_F(n-m)\bigr)}{\sin\bigl(\frac\pi{2 (L+1)}(n-m)\bigr)}-
\frac{\sin\bigl(k'_F(n+m)\bigr)}{\sin\bigl(\frac{\pi}{2(L+1)}(n+m)\bigr)}\right].
\ee

\begin{figure}[t]
\includegraphics[width=0.55\textwidth]{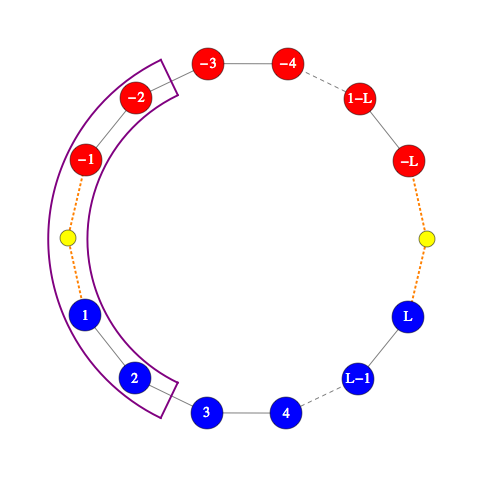}
\caption{Embedding of the open finite chain in a periodic one. The blue points correspond to the original chain,
the red ones are the mirror symmetric sites, while the two yellow points at $x=0$ and $x=L+1\equiv-L-1$ 
are ``auxiliary sites''. The picture shows a block with $\ell=2$ in the open chain corresponding to $2\ell+1=5$
in the periodic one.
}
\label{fig:emb}
\end{figure}

From this correlation matrix, it is straightforward to obtain numerical results for $S_n(\ell)$ also in finite systems. 
This analysis has been already done with considerable numerical 
accuracy in Refs. \cite{lsca-06,cv-10b} for the Von Neumann entanglement entropy and in Ref. \cite{cv-10} for general $n$.
However, the accurate results for the amplitudes of the corrections to the scaling were not compared 
with the theoretical predictions not available at that time.

\begin{figure}[t]
\includegraphics[width=0.55\textwidth]{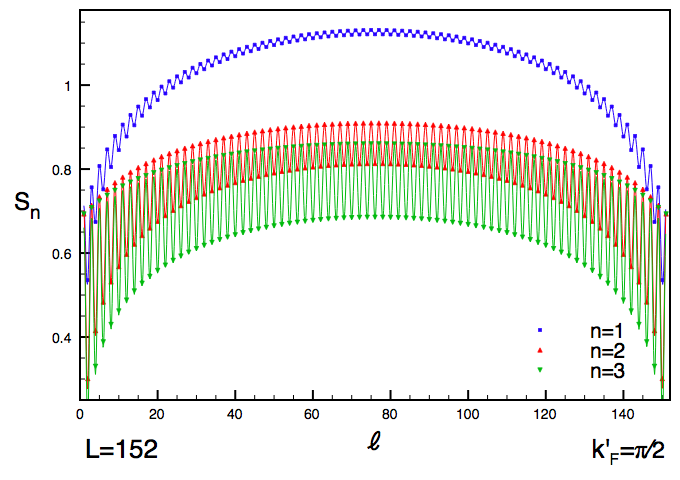}
\includegraphics[width=0.55\textwidth]{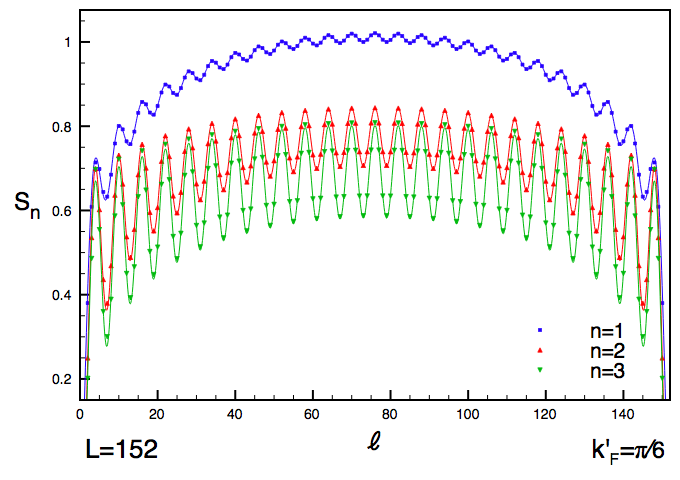}
\includegraphics[width=0.55\textwidth]{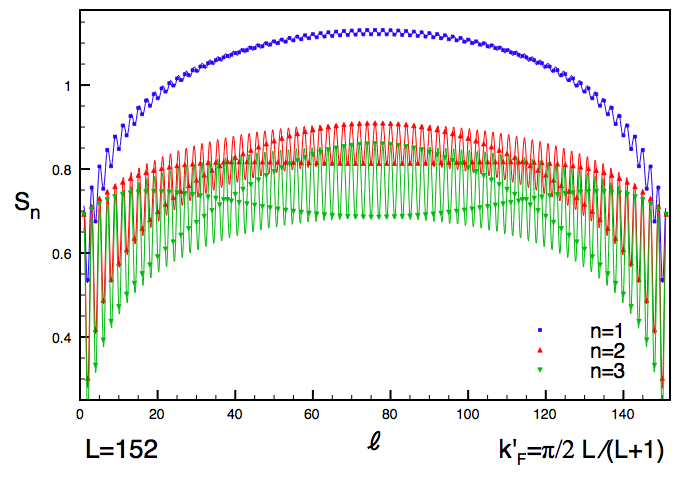}
\includegraphics[width=0.55\textwidth]{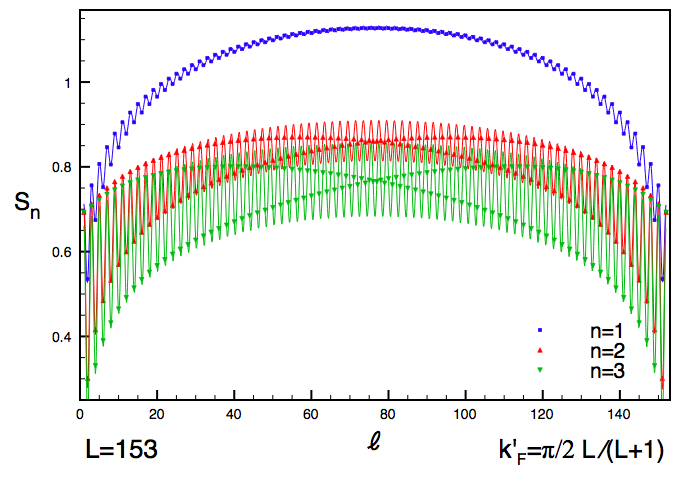}
\caption{$S_n(\ell)$ in a finite XX chain of length $L$ for $n=1,2,3$ and different values of $k_F$.
The exact numerical results for any $\ell$ are compared with the asymptotic formula including the 
first correction to the scaling. 
The excellent agreement  confirms the effectiveness of the finite-size scaling ansatz Eq. (\ref{SnFS}). 
}
\label{fig:finiteL}
\end{figure}

The modification of the leading term in Eq. (\ref{intro}) for finite systems is provided by conformal field theory \cite{cc-04}: 
the R\'enyi entropies are given by Eq. (\ref{intro}) where the length of the subsystem $\ell$ is substituted 
by the chord distance $L/\pi \sin(\pi \ell/L)$. 
However, we would like an expression that takes into account corrections to the scaling and 
that is accurate at order $1/\ell$, while we keep fixed the ratio $\ell/L$.
This is beyond the predictive power of CFT, but an intuitive argument to find a proper modification of Eq. (\ref{intro})
valid in finite size can be given, leading to the expression:
\bea\label{SnFS}
\fl S_n(\ell)&=&\frac{1}{12}\left(1+\frac1n\right)\ln \left[  \frac{4(L+1)}{\pi}\sin\frac{\pi (2\ell+1)}{2(L+1)}   |\sin k'_F|\right] 
+\frac{E_{n}}{2}\nn\fl &&+
 \frac{2\sin [k'_F (2\ell+1)]}{1-n}
\Bigl[\frac{4(L+1)}{\pi}\sin\frac{\pi (2\ell+1)}{2(L+1)}|\sin k'_F|\Bigr]^{-1/n}\frac{\Gamma(\frac{1}{2}+\frac{1}{2n})}{\Gamma(\frac{1}{2}-\frac{1}{2n})}\,.
\eea
The argument proceeds as follows. While we take the continuum limit from the spin-chain to the CFT, there is a well-known 
arbitrariness on the exact correspondence between the lattice sites and the coordinate on the continuum space. 
While for PBC, translational invariance guarantees that we can start the lattice in an arbitrary point, this is no longer true 
in the presence of boundaries. 
For a semi-infinite system,  the exact result (\ref{intro}) suggests that the first site of the chain should be placed 
at position $x=1$ in the continuum theory. 
Indeed,  when building the mirror image (as usually done in boundary CFT),
we have a mirror chain starting from $-1$ going up to $-\infty$. 
This implies that an ``auxiliary site'' should be introduced at $x=0$. 
In this way, we have an infinite chain with a block of length $2\ell+1$, exactly as Eq. (\ref{intro}) suggests. 
When we move to a finite chain of length $L$, the mirror construction is graphically depicted in Fig. \ref{fig:emb}.
We clearly have to add another auxiliary site at the other boundary to embed the open chain in a periodic one.
The resulting length of the periodic chain is $2(L+1)$.   
Thus this argument suggests that from the semi-infinite formula (\ref{intro}), 
we can obtain a finite-size ansatz  by replacing $2\ell+1$ with the modified chord length 
$\frac{2(L+1)}{\pi}\sin\frac{\pi (2\ell+1)}{2(L+1)}$.
In doing so, we should also keep in mind that the prefactor of the correction 
$\sin[k'_F (2\ell+1)]$ is not scaling (as for PBC \cite{ccen-10}) 
and there $\ell$ should be left unchanged.
All these ingredients lead to Eq. (\ref{SnFS}).

In Fig. \ref{fig:finiteL}, we report numerical calculated $S_n(\ell)$ for finite systems of different lengths and for 
different values of $k_F$. In all cases, for $n=1,2,3$, when the first correction
describes accurately the numerics for semi-infinite systems, we found perfect agreement between analytical and numerical 
results, confirming the validity of the non-rigorous argument reported above.
Notice that when $k'_F$ is not a simple number, as in the two graphics in the bottom of Fig. \ref{fig:finiteL}, 
the numerical data (points) show apparently strange periodicity. When the asymptotic exact forms are plotted (continuous lines),
it is clear that the periodicity is the correct one and the previous effect is only due to the the value of $k_F$. 

Notice that at $n=1$, the unusual correction in $\ell^{-1/n}$ and the analytic one coming from expanding  $2\ell+1$ in the leading term 
are of the same order $1/\ell$ and they have been disentangled in the past  \cite{lsca-06,cv-10b} only because 
one is oscillating and the other is not.
The presence of the non-oscillating $1/L$ term has been firstly observed and its analytic value guessed in Ref. \cite{lsca-06} 
for $n=1$.
For general $n$, its form has been correctly guessed in Ref. \cite{cv-10}. 
The oscillatory behavior has also been firstly described in Ref. \cite{lsca-06} for $n=1$, successively its amplitude has been guessed 
in Ref. \cite{cv-10b}. 
We provided an analytical proof of  this sequence of numerical guesses and we gave first analytical expressions for the 
oscillating corrections to $S_n(\ell)$ for general $n$, that was too complicated to be guessed. 

In Ref. \cite{cv-10}, for half-filling and even $L$, on the basis of numerical results, 
the R\'enyi entropies have been parametrized as 
$$
S_n= \frac{1}{12}\Bigr(1+\frac1n\Bigl)
\left[\ln [(L+1)\cos \pi X]+e_n-(-1)^{\ell} b_n {[L \cos\pi X]^{-1/n}}\right]\,,
$$
with $X=(L/2-\ell)/(L+1)$.
The numerical values $b_2\sim 4.79256$ and $b_3 \sim4.19726$ have been reported \cite{cv-10}. 
Eq. (\ref{SnFS}) is fully compatible with this expression and predicts  the  value of the amplitude $b_n$ for any $n$. 
The numerical estimates for $b_n$ in Ref. \cite{cv-10} coincide with the exact  values, 
showing the  high numerical accuracy of  Ref. \cite{cv-10}. 
Away from half-filling, in Refs. \cite{cv-10,cv-10b}, in order to force a finite-size (or finite-trap) scaling, a new parameter has been 
introduced to take into account the non-scaling term $\sin[k'_F (2\ell+1)]$, that is implicit in our definition of $k_F$.
The two representations are equivalent, but we find Eq. (\ref{SnFS}) clearer.

\section{Block disconnected from the boundary} 
\label{sec:disc}

We consider in this section the R\'enyi entanglement entropies for a block disconnected from the boundary.
The numerical calculations are straightforward and can be done exactly in the same way as before just by considering 
the correlation matrix $C_{nm}$ starting from a spin different from the first. 
We stress once again that the calculation of the spin entanglement can be done simply in terms of fermions, as
a peculiarity of the chain with open boundary conditions. 
In the case of two intervals in a PBC chain, the entanglement of spins and fermions are known to be different \cite{atc-09,ip-09} and 
also other boundary conditions are generally expected to make spins and fermions inequivalent, so that 
the calculations should be done  following the general method to tackle with the Jordan-Wigner string introduced in Ref. \cite{fc-10}.

\begin{figure}[t]
\includegraphics[width=0.7\textwidth]{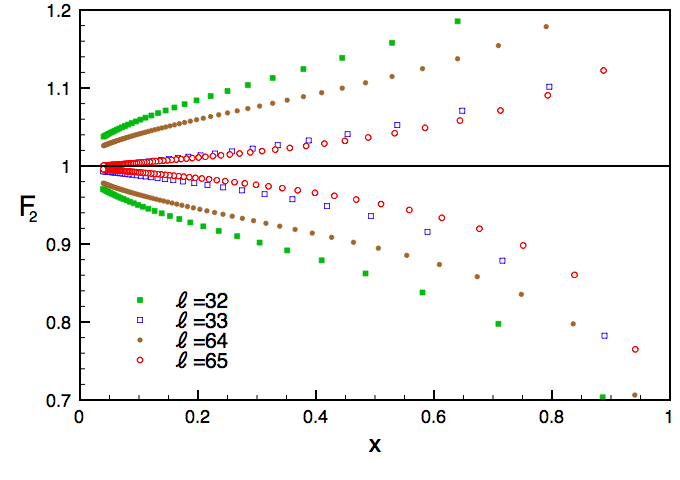}
\caption{The function $F_2^{\rm lat}(x)$ in a semi-infinite chain for different lengths of the block $\ell$. 
Finite $\ell$ corrections are large.
}
\label{fig:sta}
\end{figure}

For simplicity we will consider only systems in the thermodynamic limit (i.e. semi-infinite) and at half filling ($k_F=\pi/2$), 
but the results are very general. 
In Fig. \ref{fig:sta}, we report the function $F_2^{\rm lat}(x)$ obtained dividing ${\rm Tr}\rho_A^2$ 
by the scaling factor in Eq. (\ref{Fn}), i.e.
\be
F_2^{\rm lat}(x)\equiv \frac{\Tr\rho_A^2}{{\tilde c}_n^2 \left( \frac{(2\ell_0 +\ell)^2}{\ell^2 4\ell_0(\ell+\ell_0)}\right)^{c/12(n-1/n)}}
\ee
where $x$ is the 4-point ratio in Eq. (\ref{4pR}). 
Global conformal invariance implies that $F_2(x)$ is a function only of $x$, while in the figure we clearly see different curves for 
different $\ell$. 
As usual \cite{fps-09,atc-09,fc-10}, these differences are due to finite $\ell$ and $\ell_0$ effects that are  severe. 
In Ref. \cite{cc-10}, it has been shown that finite $\ell$ corrections are generically of the same form (i.e. 
governed by the same unusual exponent) independently of the number of blocks (that in the present case generalizes to its 
location).
Notice that the data in Fig. \ref{fig:sta} are not asymptotic also because the various curves do not have 
the conformal symmetry $x\to\ 1-x$ \cite{fps-09,cct-09} (again this is rather common \cite{fps-09,atc-09,fc-10} for finite $\ell$).

\begin{figure}[t]
\includegraphics[width=0.5\textwidth]{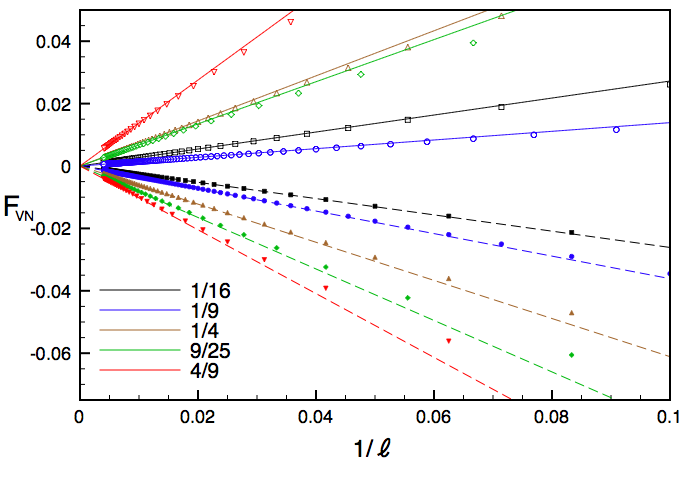}
\includegraphics[width=0.5\textwidth]{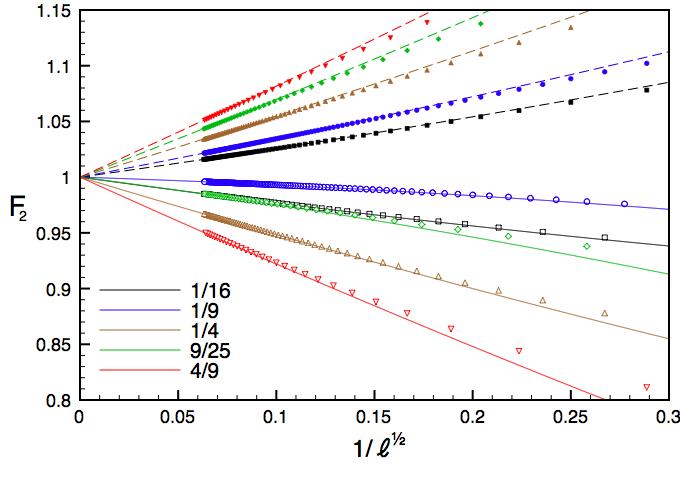}
\includegraphics[width=0.5\textwidth]{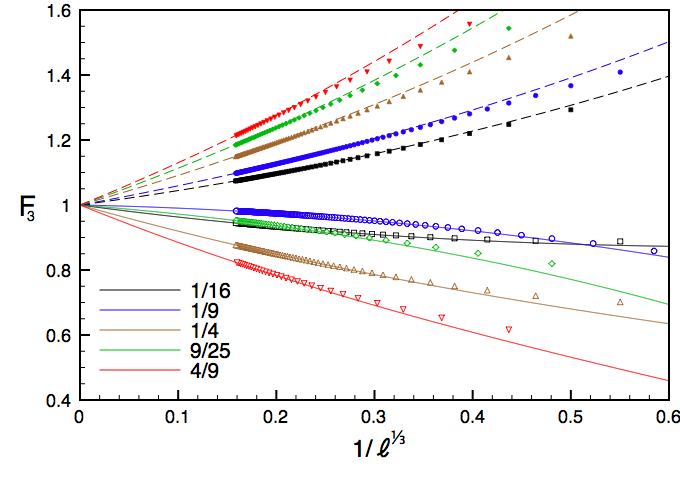}
\includegraphics[width=0.5\textwidth]{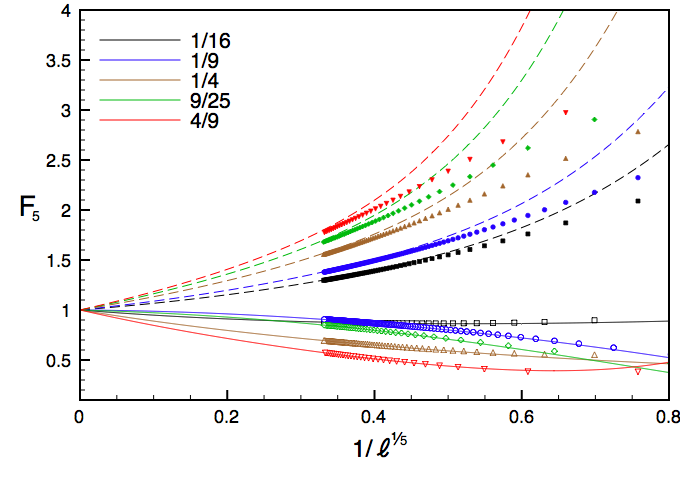}
\caption{Scaling of the finite $\ell$ corrections for $S_n(\ell)$ vs $\ell^{-1/n}$ for $n=1,2,3,5$. 
Leading corrections are of the expected form 
$\propto \ell^{-1/n}$ and the extrapolation to $\ell\to\infty$ clearly gives $F_n(x)=1$ identically. 
For large values of $n$, subleading corrections $\propto\ell^{-q/n}$ must be included.
}
\label{fig:corrFn}
\end{figure}

We can now proceed to the calculation of the asymptotic value of the function $F_n(x)$ for various $n$, by using the 
fact that corrections to the scaling are of the form $\ell^{-1/n}$. 
In Fig. \ref{fig:corrFn}, we report for $n=1,2,3,5$ the function $F^{\rm lat}_n(x)$ at fixed $x$,
obtained numerically as explained above, as function of $\ell^{-1/n}$. 
For large enough $\ell$, the points are aligned on straight lines, confirming the correctness of the finite $\ell$ scaling (increasing 
$n$, more corrections of the form $\ell^{-q/n}$ must be included to reproduce the numerical data, as obvious). 
This allows to extrapolate to $\ell\to\infty$. The result is evident from Fig. \ref{fig:corrFn}:
\be
F_n(x)=1\,,
\ee
identically. 
At first,  this can seem very strange when compared with the complicated functions found for two intervals in the XX model 
with PBC \cite{fps-09,cct-09,fc-10}.
The simplicity of this result is due to the fact that (in the present case) spins and fermions are equivalent.
For PBC free-fermions, the explicit computation shows $F_n(x)=1$ \cite{ch-v}  (as confirmed in some numerical works \cite{num}). 
This result straightforwardly generalizes to OBC free-fermions \cite{ch-rev}. 
The main peculiarity of OBC is not  $F_n(x)=1$, but  the equivalence of fermions and spins.

\section{Conclusions}
\label{sec:concl}

In this manuscript we provided a number of exact results for the asymptotic scaling of the R\'enyi entanglement 
entropies in open XX spin-chains.
Schematically our results can be summarized as follows.
\begin{itemize}
\item In the case of a semi-infinite system and a block starting from the boundary, we derive rigorously the asymptotic behavior 
for large block sizes on the basis of a recent mathematical theorem for the determinant of Toeplitz plus Hankel matrices. 
This is given by Eq. (\ref{Snobcl}).
\item To obtain the corrections to the scaling to the asymptotic behavior, we conjecture a generalized Fisher-Hartwig form for these
determinants. Eq. (\ref{main}) gives the exact asymptotic behavior of $S_n(\ell)$ at order $o(\ell^{-1})$ for any $n$. 
Eq. (\ref{Sinfty}) generalizes the result for $n=\infty$, i.e. for the largest eigenvalues of the reduced density matrix.
\item By combining these results with conformal field theory arguments, we derive exact expressions also in finite chains with open boundary conditions in Eq. (\ref{SnFS}).
\item In the case of block detached from the boundary, we  again use CFT  to derive an exact expression 
for the asymptotic $S_n$ given by Eq. (\ref{Fn}) with $F_n(x)=1$. 
\end{itemize}
All these results have been checked against exact numerical computations.

For infinite chains, in Ref. \cite{ce-10}, using a combination of methods based on the generalized Fisher-Hartwig conjecture 
and a recurrence relation connected to the Painlev\'e VI differential equation, the corrections to $S_n(\ell)$ have been 
obtained up to order $O(\ell^{-3})$. To the best of our knowledge the recurrence relation obtained in Ref. \cite{fw-05} for an infinite 
system has not been generalized to a semi-infinite one, althought a random matrix description of these chains exists \cite{km-05,f2}.

It would be also very interesting to consider the exact calculation of asymptotic and subleading terms in the R\'enyi 
entropies for different boundary conditions and the generalization to the XY model on the lines of Refs. \cite{p-04,ijk-05}.
Finally, the accurate numerical results of Refs. \cite{cv-10,cv-10b} show a very similar structure also for the case of a
general confining (trapping) polynomial potential and it is natural to wonder whether exact results can be obtained also in those cases.

\ack
We thank Fabian Essler and Ettore Vicari  for helpful discussions.

\section*{References}

\end{document}